\documentclass[aps,prl,twocolumn,superscriptaddress,showpacs]{revtex4}
\usepackage{amsmath}
\usepackage{dcolumn}
\usepackage{epsfig}
\usepackage{graphicx}
\usepackage{latexsym}

\begin{document}



\title{STM imaging of a bound state along a step on the surface of\\ the topological insulator Bi$_2$Te$_3$  }

\author{Zhanybek Alpichshev}
\affiliation{Stanford Institute for Materials and Energy Sciences, SLAC National Accelerator Laboratory, 2575 Sand Hill Road, Menlo Park, CA 94025}
\affiliation{Geballe Laboratory for Advanced Materials, Stanford University, Stanford, CA, 94305}
\affiliation{Department of Physics, Stanford University, Stanford, CA 94305}
\author{J. G. Analytis}
\affiliation{Stanford Institute for Materials and Energy Sciences, SLAC National Accelerator Laboratory, 2575 Sand Hill Road, Menlo Park, CA 94025}
\affiliation{Geballe Laboratory for Advanced Materials, Stanford University, Stanford, CA, 94305}
\author{J.-H. Chu}
\affiliation{Stanford Institute for Materials and Energy Sciences, SLAC National Accelerator Laboratory, 2575 Sand Hill Road, Menlo Park, CA 94025}
\affiliation{Geballe Laboratory for Advanced Materials, Stanford University, Stanford, CA, 94305}
\author{I.R. Fisher}
\affiliation{Stanford Institute for Materials and Energy Sciences, SLAC National Accelerator Laboratory, 2575 Sand Hill Road, Menlo Park, CA 94025}
\affiliation{Geballe Laboratory for Advanced Materials, Stanford University, Stanford, CA, 94305}
\affiliation{Department of Applied Physics, Stanford University, Stanford, CA 94305}
\author{A. Kapitulnik}
\affiliation{Stanford Institute for Materials and Energy Sciences, SLAC National Accelerator Laboratory, 2575 Sand Hill Road, Menlo Park, CA 94025}
\affiliation{Geballe Laboratory for Advanced Materials, Stanford University, Stanford, CA, 94305}
\affiliation{Department of Physics, Stanford University, Stanford, CA 94305}
\affiliation{Department of Applied Physics, Stanford University, Stanford, CA 94305}


\date{\today}

\begin{abstract}
Detailed study of the LDOS associated with the surface-state-band near a step-edge of the strong topological-insulator Bi$_2$Te$_3$ , reveal a one-dimensional bound state that runs parallel to the step-edge and is bound to it at some characteristic distance.  This bound state is clearly observed in the bulk gap region, while it becomes entangled with the oscillations of the warped surface band at high energy \cite{alpichshev}, and with the valence band states near the Dirac point.  We obtain excellent fits to theoretical predictions \cite{theory} that properly incorporate the  three-dimensional nature of the problem to the surface state \cite{hzhang}.  Fitting the data at different energies, we can recalculate the LDOS  originating from the Dirac band without the contribution of the bulk bands or incoherent tunneling effects.
\end{abstract}


\maketitle

Bi$_2$Te$_3$ and  Bi$_2$Se$_3$ have been argued recently to be three-dimensional (3D) topological insulators (TI) \cite{pt,fu,qi1,hsieh1}, exhibiting a bulk gap and a single, non-degenerate Dirac fermion surface band topologically protected by time-reversal symmetry \cite{hzhang}.  Subsequent angle resolved photoemission spectroscopy (ARPES) \cite{chen,hsieh2}, and by scanning tunneling microscopy (STM) and spectroscopy (STS) \cite{roushan,alpichshev,zhang} experiments  confirmed that prediction. Focusing on Bi$_2$Te$_3$, ARPES revealed that with appropriate hole-doping, the Fermi level could be tuned to intersect only the surface states, indicating fully gapped bulk states as is expected from a three-dimensional TI \cite{chen,hsieh2}. Complementing ARPES, STM and STS studies emphasizing quasi particle scattering interference from impurities and macroscopic defects,  demonstrated the ``protected" nature of the surface band  \cite{roushan,alpichshev,zhang}.

While protection of the surface state is guaranteed within a simple Dirac band, warping effects can produce special nesting wavevectors that allow for electronic-waves within the surface state band (SSB) \cite{fu1}. These effects have been observed experimentally in the upper part of the SSB by Alpichshev {\it et al.} \cite{alpichshev}. In that paper we reported STS  studies on high-quality Bi$_2$Te$_3$ crystals which exhibit oscillations of LDOS near a step-edge.  While within the main part of the SSB oscillations were shown to be strongly damped, supporting the hypothesis of topological protection, at higher energies, as the SSB becomes concave (as is observed by complementary ARPES data), oscillations appear which disperse with a particular wave-vector corresponding to the allowed spin states of the warped constant-energy contour \cite{fu} .

In this paper we present a more detailed study of the LDOS associated with the SSB near a step-edge,  and report on the observation of what is most probably a one-dimensional (1D)  state bound onto it.  This bound state is clearly observed in the bulk gap region, while it becomes entangled with the oscillations of the warped SSB at high energy and with the valence band states near near the Dirac point.  We successfully fit the data to theoretical predictions \cite{theory} that properly incorporate the  three-dimensional nature of the problem (outlined in \cite{hzhang})  to the surface state .  Fitting the data at different energies, we can recalculate the LDOS  originating from the Dirac band without the contribution of the bulk bands or incoherent tunneling effects. The excellent agreement between the experimental data and simple theory is another testimony to the simple, yet robust nature of the surface state in this model system of TI.

For the present study we used Cd doped single crystals of Bi$_2$Te$_3$ identical to those used in our earlier study of the warped surface band \cite{alpichshev}. Nominal doping levels of up to 1$\%$ for Cd were incorporated to compensate n-type doping from vacancy and anti-site defects that are common in the Bi$_2$Te$_3$ system.  Actual doping was determined separately using chemical and Hall-effect methods and were shown by ARPES \cite{chen} to be in excellent agreement with the relative position of the Dirac point with respect to the Fermi energy. For the data described in this paper the Dirac point was located near $\sim$300 meV.

\begin{figure}[h]
\includegraphics[width=1.0 \columnwidth]{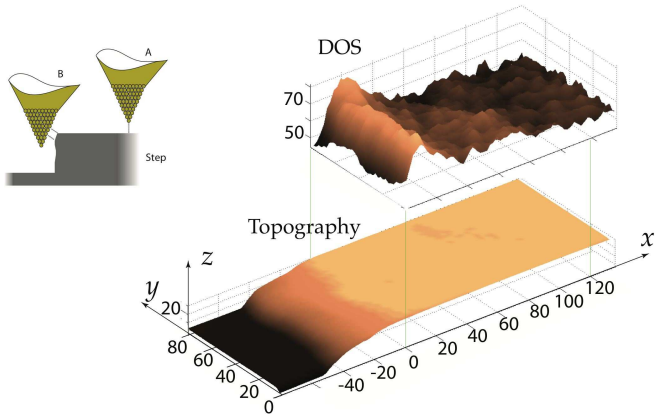}
\caption{(color online) bottom: Topography of the step. The total height of the step $\sim$30 $\AA$ corresponding to one unit cell;  top: DOS at $E=-160$ meV in the plateau region away from the step showing a bound state about 15 $\AA$ away from the step. We also show in a cartoon the tip-step interaction, showing that pure surface LDOS can be taken on the plateau away from the step (A), but due to the strong decrease in height to the left of the step, rounding of the topography must occur (B).}
\label{topo}
\end{figure}

Samples were cleaved in vacuum of better than $5\times 10^{-10}$ Torr, and quickly lowered to the $\sim$9 K section of the microscope, where cryo-pumping ensures that the surface remains free from adsorbates for weeks or longer.  Topography scans were taken at several bias voltages and setpoint currents (usually 200mV and 100pA).   A thorough discussion of the important features of the data that was collected from many topography and spectroscopy scans can be found in \cite{alpichshev}, here we concentrate on scans in the vicinity of a step-edge defect that was obtained in the process of cleaving the crystal. The measured thickness of the step is $\sim$30.5 $\AA$ (in excellent agreement with the thickness of one unit cell \cite{bite}), and we concentrate on the region away from the step as is shown in Fig.~\ref{topo}. We mark the beginning of the step which runs along the [100] direction with $x=0$ (this choice is arbitrary to within a few $\AA$ because of the roughness of the step).

Above the topography, we also show in Fig.~\ref{topo} the LDOS of that region for a bias voltage -160 meV (in the middle of the SSB) which  lies in the gap between the bulk conduction band (BCB) and the bulk valence band (BVB). While the LDOS is rather flat and featureless in most of the scanned area, a pronounced LDOS-peak is observed next to the step with constant strength along the step.  We identify this peak as a bound state and study it further as a function of energy.

As described in ref.~\cite{alpichshev}, at high enough energies, away from the Dirac point and in the region of warped surface band contour we observe oscillations of the LDOS that can be fit with the expression: $\rho(x) \propto A(E){\rm sin}(2kx+\phi)/x$  (amplitude $A$, and a scattering potential phase-shift  $\phi$).  In this region LDOS demonstrates only simple and naturally expected deviation from the simple asymptotic expression around the origin. 

\begin{figure}[h]
\includegraphics[width=1.0 \columnwidth]{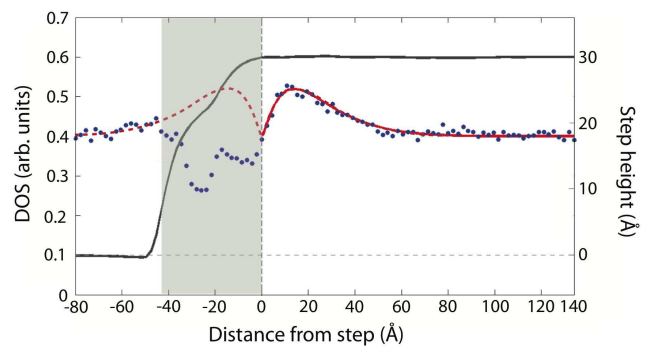}
\caption{(color online) LDOS as a function of position from the step at $E=-190$ meV (in the region of the SSB that is in the bulk gap) shown together with a fit to Eqn.~\ref{result} (solid line).  Also shown is a cross-sectional profile of the step topography. In the shaded region the tip interacts with the side of the step and its lower part (see Fig.~\ref{topo}), thus cannot be used for a fit. }
\label{model}
\end{figure}

As the energy is lowered, and we enter the region in which oscillations are strongly damped (below $\sim -100$meV), a new peak seems to emerge at a distance $\sim 15 \AA$ from the step. For example, we show in Fig.~\ref{model} raw LDOS data as a function of distance from the step (zero marks the position of the step)  in both direction, for bias energy of $-190$ meV, inside the bulk gap. For reference we also show in the same figure a cross section of the topography of the step. Note that due to the height of the step ( 30 $\AA$), when the tip drops down to keep the distance to the surface constant, it ``probes" the edge of the step which results in a rounded region that extends to $-50 \AA$. We estimate that the actual step drops down to the lower region rather steeply around zero.  To the right of the step the peak at $\sim$15 $\AA$ is the only feature that exists. The solid line shows a fit to our model \cite{theory}, as will be discussed below. We note that we can impose the mirror-image of the fit to the ``behind the step" region. The excellent agreement that is observed beyond the region in which the tip did not clear the step is a direct proof of the two-dimensional nature of the surface band that beyond the exponentially decaying effect of the bound state, it ``does not know" about the defect as if there is one, uninterrupted surface that wraps the crystal.

\begin{figure}[h]
\includegraphics[width=1.0 \columnwidth]{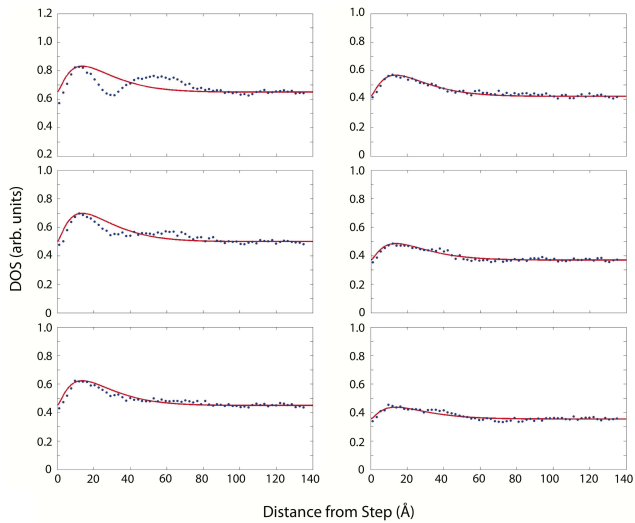}
\caption{(color online) LDOS oscillations as a function of the distance from the step for several energies in the range of the SSB that is inside the bulk gap \cite{alpichshev}. solid lines are fits to Eqn.~\ref{result} (see text).}
\label{bs}
\end{figure}

Fig.~\ref{bs} shows the raw data of LDOS near the step with similar fits to Fig.~\ref{model}.  The result at $-115$ meV seems to be a smooth progression from higher energies, if we subtract the oscillatory part, where by $\sim -150$ meV the peak stops changing shape and it can be fitted with our model with no residual.

Despite the robustness of the experimentally observed peak, neither a clear prediction, nor possible explanation exists to date to account for this phenomenon. We attribute this deficiency to the  pure-2D theoretical framework  which all of the previous studies used.  The results of these calculations (e.g. \cite{biswas1}) show that any features obtained are necessarily dispersive with energy while the peak in LDOS we report on is clearly not, which unambiguously implies some details are missing. 

While the initial Hamiltonian derived from the band structure of Bi$_2$Te$_3$ did include all bands \cite{hzhang}, all the calculations of impurity scattering or scattering from macroscopic defects used the truncated two-dimensional  $\bf k \times \sigma$ effective surface Hamiltonian first introduced by Zhang {\it et al.} \cite{hzhang}. At most, higher order surface terms were introduced to account for the hexagonal warping effects \cite{fu1}. Here we claim that by using a pure 2D Hamiltonian, one  inherently assumes that $k_x$ and $k_y$ are ``good" quantum numbers, which may be a wrong starting point in the case of a strong perturbation such as a step-edge.  When  the local curvature of the surface of a topological insulator is much less than the characteristic length-scale of the problem $\xi \sim v_F/\Delta_{bulk}$ ($\Delta_{bulk}$ is the bulk gap), one cannot assume that the surface state will be just a solution of the conventional ``$\textbf{k} \times {\bf \sigma}$" surface Hamiltonian,  wrapping the new curved surface. Instead, effects due to the gradients near the corrugated surface induce bulk interference which become relevant. Therefore, a coherent treatment of such surface defect should start with the appropriate 3D Hamiltonian such as the one derived by Zhang \textit{et al. } \cite{hzhang} (we choose the units such that $\hbar=1$):

\begin{multline}
\widehat{\mathcal {H}}\psi (\textbf{r})= \big\{M(\textbf{r}) \textbf{I}_{2\times2} \otimes \tau_z\\ + v(\hat{k}_x \sigma_x \otimes \tau_x + \hat{k}_y \sigma_y \otimes \tau_x) + v_z(\hat{k}_z \sigma_z \otimes \tau_x)\\ + \mathcal{O}(k^2)\big\} \psi (\textbf{r})=E\psi (\textbf{r})
\label{sc}
\end{multline}
where $\sigma_i$ and $\tau_i$   are Pauli matrices in  spin and orbital space respectively with $i=x;y;z$, and $\textbf{I}_{2 \times 2}$ is the $2 \times 2$ identity matrix. $v$, $v_z$, and $M(\textbf{r})$ are material-specific parameters. In particular, the ``mass term" $M(\textbf{r})$ represents the asymmetry between the inside and outside the bulk of the TI material and has units of energy in our notations. Asymptotically $M(\textbf{r}) =-M_{in}<0$   inside,  and $M(\textbf{r}) =M_{out}>0$ outside, while near the step $M(\textbf{r})$ varies rapidly to match the asymptotic behavior. It is easy to see that from the topology point of view, it is enough to set  $M(\textbf{r})=+M$ in vacuum and $M(\textbf{r})=-M$ in the TI bulk in order to create a boundary that separates the two distinct topological states . Further, to localize the effect of the mass twist to the TI surface the Eq.~\ref{sc} can be squired. The result is a new Schr\"odinger-like equation (for convenience, the $z$-axis was re-scaled such that $\tilde{z} = z(v/v_z)$) for which it can be demonstrated that the LDOS obtained from it is directly related to the LDOS of the original unsquared Hamiltonian. 
 
\begin{multline}
\big\{ \left[ M^2 - v^2 \left(  \partial_x^2 + \partial_y^2 +  \partial_{\tilde{z}}^2 \right) \right] \textbf{I}_{4 \times 4}\\ - v(\nabla M(\textbf{r}) \cdot \sigma ) \otimes \tau_y \big\}\psi (\textbf{r})= E^2 \psi (\textbf{r}).
\label{squared}
\end{multline}

\noindent However, despite the fact that the effect of the step is now separated into the gradient term, Eq.~\ref{squared} is still difficult to solve due to the unique spin structure associated with the eigenstates. However, one may argue that since at least one of the edges of the step (the one perpendicular to the cleavage surface) is very rough, it may serve as a termination line for the spin state. 

Close examination of the geometry of the step as shown in Fig.~\ref{topo} indicates a bound surface state that is $\sim15 \AA$ away from the $x=0$ axis, and spreads along the  $y$-axis. The step, which is a line defect in the $x$-$y$ plane, is created by an intersection of a surface perpendicular to the $z$-axis with a surface perpendicular to the $x$-axis. In such a formulation $k_y$ continues to be a good quantum number in the plane. Therefore, setting $k_y=0$, we make an approximation in which we assume that we can ignore the spin structure on the rough surface of the step, hence obtaining the equation:

\begin{equation}
(M^2-v^2\nabla^2)|\psi|-v|\nabla M| |\psi|=0
\end{equation}

\noindent Solving to the first approximation this equation without the spin variables as discussed above yields:

\begin{equation}
\frac{|\psi (x)|}{|\psi_0|}= \frac{1}{2}+\frac{1}{2}e^{-x/\lambda}+\frac{1}{\lambda}\int_0^x \frac{K_0(x^\prime/\lambda)}{\pi} dx^\prime
\label{result}
\end{equation}

\noindent where $|\psi_0|$ is the asymptotic amplitude of the wavefunction far away from the step, $K_0(x)$ is the zero$^{th}$-order modified Bessel function of the second kind, and $\lambda = v/M \simeq \hbar v_F / \Delta_{bulk} \simeq 30 \AA$ is the characteristic length-scale for variations in the $x$-direction. 

\begin{figure}[h]
\includegraphics[width=1.0 \columnwidth]{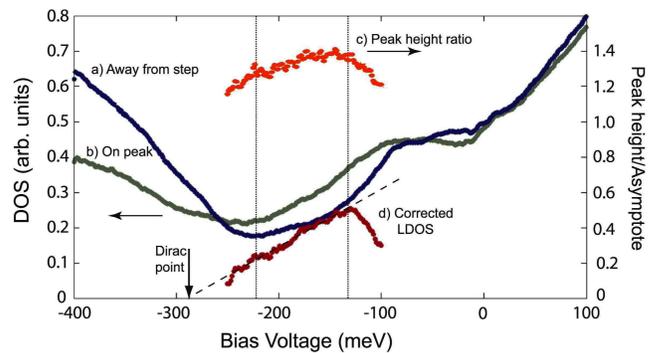}
\caption{(color online) LDOS spectra as a function of energy (left axis) away from the step (a), at the position of the maximum of the peak (b), the ratio between the peak height and the asymptotic LDOS (c), and  the extracted coherent part of LDOS (d) which is proportional to the difference between the blue and green curves (see text). Also shown is the position of the Dirac point for this crystal inferred from the analysis of LDOS oscillations and ARPES.  The region between the vertical dotted lines is within the bulk gap of the crystal.}
\label{position}
\end{figure}

An examination of the above solution shows  that $|\psi (x)|/|\psi_0|$ has a single maximum ($\psi_{max}$) near the step and then decays exponentially far away from the step. The corresponding LDOS ratio, $\alpha(E) \equiv \rho_{max}/\rho_0 \approx |\psi_{max}|^2/|\psi_0|^2 $,  is found to be approximately 1.2, and independent of energy for the wavefunction solution in Eqn.~\ref{result}. Obviously, the total density of states will be built of a combination of wavefunctions that take into account the actual roughness of the surface and the correct spin configuration. While such a calculation is beyond the scope of the present paper, we expect that the shape of the correct solution solution, i.e. a density of states that peaks close to the step and decay asymptotically far from the step, will be of the same nature \cite{theory}.  Apart from particular details, within the presented approach the structure of the peak in LDOS is determined by the bulk properties of the samples, i.e. on the energy scales of the gap size implying that in the low energy regime (close to the Dirac point) the LDOS profile should be independent of energy. This is certainly the case for the position of LDOS maximum and the shape of the profile up to a vertical shift. However it also means that the relative height $\rho_{max}/\rho_0$ of the LDOS must be independent of energy too.  Fig.~\ref{position} shows a typical LDOS spectrum away from the step, together with the ratio of peak height to asymptotic level of the LDOS as determined from the experiment as a function of bias voltage.  What it shows though is that in the exposed region of the Dirac band, the ratio $\alpha(E)$ is clearly energy dependent.  We assume that this dependence is due to an incoherent density of states $\rho_{inc}(E)$, such that $\rho_{total}(E,x)=\rho_{coh}(E,x)+\rho_{inc}(E)$.  Assuming that the coherent part of LDOS  $\rho_{coh}(E,\infty)$ behaves as explained above we can obtain from here an expression for it in terms of measurable quantities only:

\begin{equation}
\rho_{coh}(E,\infty)=\rho_{tot}(E,\infty)\cdot\left( \frac{\alpha_{exp}(E)-1}{\alpha_{max}-1}\right)
\label{ratio}
\end{equation}
 
\noindent Here $\alpha_{max}$ is the relative hight of the peak in the case the was no incoherent contibution to the LDOS. Since the latter can only decrease the height of peak, for specificity we take $\alpha_{max} = 1.35$ here, which is the maximum value of $\alpha_{exp}(E)$. The actual value for $\alpha_{max}$ should be given by the complete theory.  The resulting coherent LDOS obtained this way is plotted in Fig.\ref{position} as a function of energy. It is seen immediately that the resulting line is strikingly straight and points exactly to the position where the Dirac point is expected to be located based on electron wave analysis presented in \cite{alpichshev}, as well as ARPES data on these samples \cite{chen}. To justify that we extrapolated the validity region of Eq.\ref{ratio} quite far way from the Dirac point although it is formally applicable only in the vicinity of DP, we point out that first of all, the actual criterion was $(E/M)^2\sim(E/\Delta)^2<<1$ which is still roughly met when $(E-E_F)\approx-140 meV$; and second it is seen from Fig.~\ref{bs} that for these energies the Eqn.~\ref{result} is still providing a good fit, suggesting that the physics involved in obtaining Eqn.~\ref{result} is still valid. 

In conclusion we report on the observation of an accumulation of DOS near an atomic step on the surface of Bi$_2$Te$_3$ identified as a bound state, whose structure is found to be in nice quantitative agreement with the theory developed in ref~\cite{theory}. We also introduce a technique to extract the coherent part of the DOS, and demonstrated that unlike the total measured DOS, its shape is in accordance with the expected profile of DOS of a 2D Dirac band.

\acknowledgments
Discussions with Xiaoliang Qi, Shoucheng Zhang, and especially Srinivas Raghu are greatly appreciated.  This work was supported by the Department of Energy Grant  DE-AC02-76SF00515.

\end{document}